\documentstyle[12pt,epsf,epsfig]{article}
\begin{document}
\baselineskip 0.6cm

\newcommand{\gsim}{ \mathop{}_{\textstyle \sim}^{\textstyle >} }
\newcommand{\lsim}{ \mathop{}_{\textstyle \sim}^{\textstyle <} }
\newcommand{\vev}[1]{ \langle {#1} \rangle }
\newcommand{\EV}{ {\rm eV} }
\newcommand{\KEV}{ {\rm keV} }
\newcommand{\MEV}{ {\rm MeV} }
\newcommand{\GEV}{ {\rm GeV} }
\newcommand{\TEV}{ {\rm TeV} }
\newcommand{\barr}[1]{ \overline{{#1}} }
\newcommand{\nn}{\nonumber}
\newcommand{\ra}{\rightarrow}
\newcommand{\bino}{\tilde{\chi}}
\def\tr{\mathop{\rm tr}\nolimits}
\def\Tr{\mathop{\rm Tr}\nolimits}
\def\Re{\mathop{\rm Re}\nolimits}
\def\Im{\mathop{\rm Im}\nolimits}
\setcounter{footnote}{1}

\begin{titlepage}

\begin{flushright}
UCB-PTH-00/45\\
LBNL-47230\\
UT-955\\
\end{flushright}

\vskip 1.5cm
\begin{center}
{\large \bf Gauge Mediation Models with Neutralino Dark Matter}
\vskip 1.0cm
Yasunori Nomura$^{1,2}$ and Koshiro Suzuki$^{3}$

\vskip 0.4cm

$^{1}$ {\it Department of Physics, University of California, 
         Berkeley, CA 94720, USA}\\
$^{2}$ {\it Theoretical Physics Group, 
         Lawrence Berkeley National Laboratory, \\
         Berkeley, CA 94720, USA}\\
$^{3}$ {\it Department of Physics, University of Tokyo, 
         Tokyo 113-0033, Japan}

\vskip 1.5cm

\abstract{
We study gauge mediation models of supersymmetry breaking 
with neutralino LSP.  These models are naturally realized 
by embedding the usual four-dimensional gauge mediation models 
into a higher-dimensional spacetime such as $M^4 \times S^1 / Z_{2}$.
We calculate the relic abundance of the neutralino LSP in these models 
and show that there exist wide parameter regions where the neutralino 
LSP constitutes the dominant component of the cold dark matter.  
These regions evade constraints from collider experiments such as 
Higgs mass bounds and $b \ra s \gamma$, and also provide the value 
for the muon anomalous magnetic moment which is consistent with the 
SUSY explanation of the deviation from the standard model prediction.}

\end{center}
\end{titlepage}

\renewcommand{\thefootnote}{\arabic{footnote}}
\setcounter{footnote}{0}

%
%
%
%

\section{Introduction}

Low-energy supersymmetry (SUSY) with dynamical SUSY breaking (DSB) is
a very attractive framework for explaining the large hierarchy
between the electroweak and the Planck scales.
One of the interesting features of this framework is that the lightest 
supersymmetric particle (LSP) is completely stable in models with 
$R$ parity, which is well motivated by the stability of the proton.
In conventional hidden-sector SUSY-breaking scenario, the LSP is believed 
to be the lightest neutralino, providing a good candidate for the cold 
dark matter in the universe \cite{LSP-DM}.  It is, indeed, suggested 
that the presence of a stable particle of weak-scale mass with 
electroweak interactions is a crucial ingredient for a natural solution 
to the cosmic coincidence problems \cite{AHKM}.  In fact, in view of its 
attractiveness, the possibility of neutralino dark matter has been 
extensively studied within the context of the hidden-sector scenario 
such as constrained version of the minimal SUSY standard model (MSSM) 
\cite{DM_hidden-1, DM_hidden-2, DM_hidden-3, DM_hidden-4, EFOS}.

However, it is known that the hidden-sector scenario has no firm 
theoretical foundation.  In this scenario, the K\"ahler potential is 
assumed to take a canonical form, but it can hardly be justified 
in supergravity.  In supergravity, we expect the presence of the 
following non-renormalizable interactions in the K\"ahler potential:
\begin{equation}
  K = \frac{\eta_{ij}}{M_G^2} Z^{\dagger} Z\, Q_i^{\dagger} Q_j,
\label{Kahler-FCNC}
\end{equation}
which cannot be forbidden by any symmetry of the theories.  Here, $Z$ 
is the superfield responsible for the SUSY breaking and $Q_i$ represents
generic standard-model quark and lepton superfields with $i,j=1,\cdots,3$ 
being flavor indices; $M_G \simeq 2.4 \times 10^{18}~\GEV$ is the 
gravitational scale and $\eta_{ij}$ are constants of order one.  
These operators invalidate the assumption of the hidden-sector 
scenario, say the boundary condition of the constrained MSSM.
Therefore, in the absence of any specific realization of the 
hidden-sector scenario, the mass spectrum used in most analyses of 
neutralino dark matter must be regarded as an ad hoc working
hypothesis.\footnote{
The special case that $m_0 = m_{1/2} = A/3$ can be realized in 
five-dimensional theories with all the standard-model fields 
living in the bulk \cite{BHN2}.}
Moreover, since there is no reason for the above operators to be flavor 
universal, they generically induce too much flavor-changing neutral 
currents (FCNC's) in the SUSY standard-model sector.  In fact, 
suppressing FCNC is one of the most important issues in the SUSY 
standard model.

Gauge-mediated SUSY breaking (GMSB) \cite{org-GMSB, min-GMSB, GMSB} 
provides an elegant solution to this SUSY FCNC problem.
In GMSB models, the squark and slepton masses are generated by the 
standard-model gauge interactions and are automatically flavor universal.
On the other hand, in most GMSB models, the SUSY-breaking scale is 
supposed to be much lower than that in the conventional hidden-sector 
scenario. As a result, the gravitino mass $m_{3/2}$ becomes much
smaller than the electroweak scale \cite{GMSB}, and the neutralino is
no longer the LSP.  The reason is again due to the non-renormalizable 
interactions Eq.~(\ref{Kahler-FCNC}).  Since these operators would 
violate flavor universality, they generate flavor non-universal pieces 
of the soft SUSY-breaking masses for the squarks and sleptons of the 
order of the gravitino mass $m_{3/2} \simeq F_Z/M_G$.  To preserve the 
success of GMSB models, these flavor non-universal pieces should 
be much smaller than the flavor universal pieces generated by the gauge 
mediation, which forces the gravitino to be much lighter than the 
electroweak scale.

The question then is whether there are explicit models which solve 
FCNC problems and also accommodate neutralino dark matter.
Notice that the above argument is entirely based on the presence of the 
non-renormalizable operators in Eq.~(\ref{Kahler-FCNC}), and it comes 
from the expectation in four-dimensional effective field theories 
that all operators consistent with the symmetries are present 
with coefficients of order one suppressed by some cut-off 
(gravitational) scale.  However, this naive expectation does not 
necessarily hold if fundamental theories are higher dimensional.
Specifically, if $Z$ and $Q_i$ fields are localized on different
$(3+1)$-dimensional branes in higher-dimensional spacetime, there is no
direct contact interaction between $Z$ and $Q_i$ fields and the 
flavor non-universal operators in Eq.~(\ref{Kahler-FCNC}) are 
exponentially suppressed by the distance of two branes \cite{bulk, RS}.
Therefore, it is possible that the situations for the hidden-sector 
and GMSB models are drastically changed by embedding these models 
into higher dimensional theories at high energy scales.
However, it turns out that the hidden sector still does not work, since 
necessary flavor universal pieces for the squark and slepton masses are 
also suppressed in this case \cite{RS}.\footnote{
One possibility is to put the standard-model gauge fields in the bulk, 
and generate the squark and slepton masses through 
renormalization-group effects below the compactification scale 
\cite{gaugino}.  We here consider the case where the standard-model 
gauge and Higgs fields are also localized on the same brane as $Q_i$'s.}
In the GMSB case, in contrast, the squark and slepton masses are 
generated by the standard-model gauge interactions.  Thus, if we 
somehow manage the separation of $Z$ and $Q_i$, it merely says 
that the gravitino mass no longer has to be smaller than the 
electroweak scale.

In fact, in a class of GMSB models \cite{min-GMSB, NTY} the sector
responsible for DSB can be fully separated from the sector that feels
standard-model gauge interactions; 
two sectors communicate with each other only through the U(1)$_m$ 
gauge interaction called messenger gauge interaction.
Therefore, it is interesting to interpret these GMSB models as low-energy
manifestations of the following brane-world scenario \cite{NY}.
All fields in the DSB sector live on the DSB brane while the messenger
and the standard-model fields are localized on our observable brane. 
The U(1)$_m$ gauge multiplet is put in the bulk, through which two 
sectors on different branes can communicate with each other.
Then, the SUSY breaking on the DSB brane is transmitted to the
observable brane only by the U(1)$_m$ gauge and gravitational 
interactions across the bulk.\footnote{
Another mechanism of transmitting SUSY breaking between two branes is
discussed in Ref.~\cite{INY}.}
The non-renormalizable operators in Eq.~(\ref{Kahler-FCNC}) are absent 
(exponentially suppressed) and flavor non-universal soft masses 
are not generated.

Let us see how the above prescription works explicitly.
It has been claimed \cite{LS, RS} that the brane separation produces 
the following no-scale type K\"ahler potential in low-energy 
four-dimensional effective field theories:\footnote{
The no-scale supergravity \cite{no-scale} adopts a specific form 
$f_D = Z + Z^{\dagger}$.
We here assume the K\"ahler potential for the $Z$ field to be of the
form $f_D = Z^{\dagger} Z + \cdots$, where the ellipsis denotes higher
order terms.}
\begin{equation}
  K = -3\log \left(1 
    - \frac{1}{3}f_O(\Phi_{\rm obs}, \Phi_{\rm obs}^{\dagger}) 
    - \frac{1}{3}f_D(\Phi_{\rm DSB}, \Phi_{\rm DSB}^{\dagger})\right).
\label{Kahler}
\end{equation}
Here, $\Phi_{\rm obs}$ and $\Phi_{\rm DSB}$ denote superfields in the
observable and DSB sectors, respectively.
With the above K\"ahler potential, all soft SUSY-breaking masses and 
$A$ terms in the observable sector vanish at the tree level in the 
limit of zero cosmological constant \cite{IKYY, RS}.
Then, the soft SUSY-breaking masses are generated only by the gauge
mediation caused by the loop effects of the bulk U(1)$_m$ gauge
interaction \cite{NY}.  On the contrary, the $\mu$ term naturally 
arises at the tree level \cite{GM} from the K\"ahler potential 
if $f_O$ contains $f_O \supset H_u H_d$, where $H_u$ and $H_d$ 
are chiral superfields of Higgs doublets.
This mechanism \cite{GM} produces the SUSY-invariant mass $\mu$ of the
order of the gravitino mass, {\it i.e.} $\mu \simeq m_{3/2}$.
Therefore, if we choose the gravitino mass to be 
$m_{3/2} \simeq 100~\GEV-1~\TEV$, we can correctly reproduce the
electroweak symmetry breaking.\footnote{
With these gravitino masses the anomaly mediation \cite{RS, GLMR}
generates too small SUSY-breaking masses in the observable sector.}
Putting the U(1)$_m$ in the bulk, indeed, the transmission of the SUSY
breaking becomes necessarily weak compared with purely four-dimensional
GMSB models.  Thus, we can naturally obtain the gravitino mass 
$m_{3/2} \simeq 100~\GEV-1~\TEV$ with moderate size of extra dimensions.
In the five-dimensional model of Ref.~\cite{NY}, for instance, the
compactification length $L$ is given by 
$L \simeq (2 \times 10^{15}~\GEV)^{-1}$, which is close to the
value obtained in the scenario of Ref.~\cite{bulk}.
This provides a simple solution to the $\mu$ problem in GMSB models
with DSB and observable sectors geometrically separated in
higher-dimensional spacetime \cite{INY, NY}.

The consequence of the above brane-world GMSB models is that the
gravitino has a mass of order $100~\GEV-1~\TEV$, although the mass
spectrum of the gauginos, squarks and sleptons is the same with that of 
the usual GMSB models \cite{INY, NY}.  Thus, the lightest neutralino is 
most likely the LSP and can behave as a cold dark matter in the universe.
In this paper, we calculate the relic abundance of the neutralino LSP in
these brane-world GMSB models and show that the neutralino can actually 
be the dark matter in a wide range of the parameter space.
Since GMSB models are highly predictive, the relic abundance
$\Omega_{\tilde{\chi}}$ is calculated in terms of a few parameters: 
one parameter smaller than in the case of constrained MSSM.
Requiring that $\Omega_{\tilde{\chi}}$ is in the cosmologically favored
region, $0.1 \lsim \Omega_{\tilde{\chi}} h^2 \lsim 0.3$ ($h$ is the
present day Hubble parameter in units of 
$100~{\rm km}\,{\rm s}^{-1}\,{\rm Mpc}^{-1}$), we obtain the constraint 
on the parameter space of the models and hence masses for the 
superparticles.  We also calculate the lightest Higgs boson mass and 
the constraint from $b \rightarrow s \gamma$ process to identify the 
phenomenologically allowed region of the parameter space.
The resulting region is remarkably consistent with the SUSY
explanation of the recently reported 2.6$\sigma$ deviation 
\cite{muon} of the muon anomalous magnetic moment from the 
standard-model value.
We assume the conservation of $R$ parity throughout the paper.

\section{Gauge-Mediated SUSY Breaking Model}

The mass spectrum of the superparticles has a great influence on
the estimation of the LSP relic abundance. Since brane-world GMSB 
models have the same mass spectrum as ordinary four-dimensional
GMSB models except for the gravitino, we begin with reviewing 
the soft masses for the gauginos and sfermions in GMSB models.

In GMSB models, the superparticle mass spectrum does not depend 
on the detail of the DSB sector.  Thus, it is sufficient to simply 
consider the messenger sector which consists of $N$ pairs of 
vector-like messenger superfields $q_{i}$ and $\bar{q}_{i}$ 
($i = 1,\cdots,N$).  To preserve the gauge coupling unification, 
$q_{i}$ is supposed to form a complete grand unified theory 
(GUT) multiplet.  We assume, in this paper, that $q_{i}$ and 
$\bar{q}_{i}$ transform as ${\bf 5}$ and ${\bf 5}^{\star}$ 
representation under the SU(5)$_{\rm GUT}$.
Then, the superpotential for the messenger sector is written as
\begin{equation}
  W = \sum_{i=1}^{N} \left( 
    \lambda^{d}_{i} S d_{i} \bar{d}_{i} + 
    \lambda^{l}_{i} S l_{i} \bar{l}_{i} 
  \right),
\label{eq:mess-sp}
\end{equation}
where $(d_i, l_i) \in q_i$ and $(\bar{d}_i, \bar{l}_i) \in \bar{q}_i$.
The messenger quark multiplets $d$ and $\bar{d}$ transform as the 
right-handed down quark and its antiparticle under the standard-model 
gauge group, respectively, and the messenger lepton multiplets $l$ and 
$\bar{l}$ as the left-handed doublet lepton and its antiparticle, 
respectively.
The singlet field $S$ is a spurion superfield parameterizing the 
SUSY-breaking effect, and assumed to have vacuum expectation values 
(VEV's) in its lowest and highest components as 
$\vev{S} = M + \theta^2 F$.

The SUSY breaking effect in the messenger sector is transmitted to 
the superpartners of the standard-model particles through the 
standard-model gauge interactions.
The gaugino masses and the sfermion squared masses are generated 
at the messenger scale, $M$, by one- and two-loop diagrams, respectively.
In general, various coupling constants, $\lambda^{d}_{i}$ and 
$\lambda^{l}_{i}$, in the superpotential Eq.~(\ref{eq:mess-sp}) are 
not equal at the messenger scale; for instance, the couplings 
of the messenger quarks and leptons are not the same at the scale $M$, 
$\lambda^{d} \neq \lambda^{l}$, even if we assume they are equal 
at the GUT scale.
However, the effect of having different couplings on the superparticle 
mass spectrum does not appear at the leading order in $F/M^2$, so that 
we here take all the coupling constants in Eq.~(\ref{eq:mess-sp}) to be 
equal for simplicity and absorb it into the definition of $M$ and $F$.
Then, the gaugino masses $M_{a}$ and the sfermion squared masses 
$m_{\tilde{f}}^{2}$ are given by \cite{DGP, SP-M}:
\begin{eqnarray}
  M_{a} &=& N \frac{\alpha_{a}}{4\pi} \frac{F}{M}
  \, {\cal G}\!\left(\left| \frac{F}{M^2} \right|\right), 
\label{gaugino_mass} \\
  m_{\tilde{f}}^{2} &=& 2 N \left| \frac{F}{M} \right|^2 
  \sum_{a} \left( \frac{\alpha_{a}}{4\pi} \right)^{2} C_{a}^{\tilde{f}}
  \, {\cal F}\!\left(\left| \frac{F}{M^2} \right|\right), 
\label{sfermion_mass}
\end{eqnarray}
where $a = 1,2,3$ represents the standard-model gauge groups and 
$C^{\tilde{f}}_{a}$ is the quadratic Casimir coefficient for the 
representation each sfermion belongs to.
Here, the functions ${\cal G}$ and ${\cal F}$ are defined by
\begin{eqnarray}
  {\cal G}(x) &=& \frac{1}{x^{2}} \left[ 
    (1+x) \ln (1+x) + (1-x) \ln (1-x) \right], \\
  {\cal F}(x) &=& \frac{1+x}{x^{2}} \left[
    \ln (1+x) - 2 {\rm Li} \left( \frac{x}{1+x} \right) + 
    \frac{1}{2} {\rm Li} \left( \frac{2x}{1+x} \right)
    \right] + (x \ra -x).
\label{exact-softmass}
\end{eqnarray}
These functions have properties that ${\cal G}(x) \simeq {\cal F}(x) 
\simeq 1$ when $x \ll 1$.

To obtain the mass spectrum of superparticles, we have to evolve 
soft masses given in Eqs.~(\ref{gaugino_mass}, \ref{sfermion_mass}) 
using renormalization-group equations (RGE).
Together with the contribution from the electroweak symmetry breaking,
the masses for all the superparticles are determined.
The Higgs sector contains two more parameters $\mu$ and $\mu B$:
the SUSY-invariant ($W = \mu\, H_u H_d$) and SUSY-breaking 
(${\cal L} = \mu B\, h_u h_d$) mass terms for the Higgs doublets.
In brane-world GMSB models, both $\mu$ and $B$ are generated of the 
order of the weak scale, in contrast to the minimal model discussed 
in Ref.~\cite{BCW} where $B=0$ is assumed at the messenger scale.
The electroweak symmetry breaking condition relates these two parameters 
to $v$ and $\tan \beta$ up to the sign of $\mu$, where 
$v \equiv \sqrt{\vev{h_u}^2+\vev{h_d}^2} \simeq 175~{\rm GeV}$ 
and $\tan \beta \equiv \vev{h_u}/\vev{h_d}$.
Therefore, we end up with the following set of free parameters in our 
analyses: $N$, $F/M$, $M$, $\tan \beta$ and sgn($\mu$).

With the above GMSB mass spectrum, the correct electroweak symmetry 
breaking requires rather large value of the $\mu$ parameter, 
$\mu \gsim 2 M_2$.  This is because in GMSB models the colored 
particles are relatively heavy and, as a result, the Higgs-boson mass 
squared receives large negative contribution from the top squark 
through the top Yukawa coupling.  This fact has some important 
consequences in the present brane-world GMSB scenario.
First, since $\mu$ is of the order of the gravitino mass, the gravitino 
tends to be heavier than the lightest superpartner of the standard model.
Thus, the gravitino is not the LSP in contrast to the usual 
four-dimensional GMSB models.
According to Eqs.~(\ref{gaugino_mass}, \ref{sfermion_mass}), then, 
the lightest neutralino $\tilde{\chi}$ or the right-handed stau 
$\tilde{\tau}_{R}$ can be the LSP.  In this paper, we concentrate on 
the case where $\tilde{\chi}$ is the LSP and $\tilde{\tau}_{R}$ is the 
next to LSP (NLSP), since $\tilde{\tau}_{R}$ LSP leads to the serious 
problem of charged dark matter.  Indeed, the LSP is $\tilde{\chi}$ 
in most of the parameter space, especially when $N=1$.

Another important consequence of $\mu \gg M_2$ is that the lightest 
neutralino is almost purely composed of the bino.
This leads to a significant simplification in understanding the 
neutralino relic abundance; for instance, annihilation into $Zh$ is 
strongly suppressed due to the smallness of the Higgsino component 
in $\tilde{\chi}$.  In the next section, we list the processes 
relevant for determining the relic abundance of the lightest 
neutralino $\tilde{\chi}$, and calculate various quantities in the 
present models.

\section{Relic LSP Abundance}

The LSP is stable in $R$-parity preserving models, and hence its 
number density can decrease only through annihilation processes. 
In an expanding universe, the pair-annihilation ``freezes out'' when 
the expansion rate of the universe exceeds the interaction rate of 
the annihilation.  After the freeze out, the number density of the LSP 
per comoving volume is constant, so that some amount of relic LSP 
is left in the present universe.  Neutralino LSP abundance has been 
well studied in the context of conventional hidden-sector SUSY breaking
models \cite{DM_hidden-1, DM_hidden-2, DM_hidden-3, DM_hidden-4, EFOS}.
It is easily estimated once we could determine the relevant cross 
sections for the annihilation of the LSP.  We first briefly review 
the procedure of calculating the relic abundance.

The time evolution of the LSP number density is described by the
Boltzmann equation.  Since in GMSB models the bino and the right-handed 
stau are degenerate in some parameter region, we have to take into 
account the coannihilation effect \cite{GS, MY}.
The Boltzmann equation with the coannihilation effect can be written 
as an equation for $n = \sum_{i} n_{i}$, where $n_{i}$ are the number 
densities of the species ${i}$, and ${i}$ represents the LSP neutralino 
$\tilde{\chi}$ and the right-handed charged sleptons $\tilde{\tau}_{R}$, 
$\tilde{\tau}_{R}^{*}$, $\tilde{\mu}_{R}$, $\tilde{\mu}_{R}^{*}$, 
$\tilde{e}_{R}$ and $\tilde{e}_{R}^{*}$. 
The equation takes the following form \cite{EFOS}:
\begin{equation}
  \frac{dn}{dt} = -3Hn - \vev{\sigma_{\rm eff} v} 
  \left[ n^{2} - (n^{{\rm eq}})^{2} \right],
\label{boltzmann}
\end{equation}
where $H$ is the Hubble parameter and 
\begin{equation}
  \vev{\sigma_{\rm eff}v} = \sum_{i,j}^{} \vev{ \sigma_{ij} v }
  \frac{n_{i}^{{\rm eq}}}{n^{{\rm eq}}} 
  \frac{n_{j}^{{\rm eq}}}{n^{{\rm eq}}}.
\label{sigma_eff}
\end{equation}
Here, $n^{\rm eq}$ ($n^{\rm eq}_{i}$) is the equilibrium value of $n$
($n_{i}$) and $v$ is the relative velocity of the particles $i$ and $j$. 
The bracket denotes the thermal average and $\sigma_{ij}$ is the total 
annihilation cross section of $i+j \ra X+X'$:
\begin{equation}
  \sigma_{ij} = \sum_{X,X'} \sigma( i+j \ra X+X' ),
\end{equation}
where $X$ and $X'$ represent possible standard-model particles.

There are $7 \times 7$ cross sections ($\sigma_{ij}$'s) in
Eq.~(\ref{sigma_eff}), but most of them are not independent. The
independent cross sections are listed in Table~{\ref{tab.coanh}}, where 
we have shown only relevant final states which are kinematically 
accessible and have non-negligible cross sections.
Furthermore, there are some simplifications coming from the GMSB 
mass spectrum.  For $\tilde{\chi} \tilde{\chi} \ra f \bar{f}$, for 
example, the cross section is dominated by $l_{R} \bar{l}_{R}$ final 
states since the right-handed sleptons are much lighter than the other 
sfermions and have the largest value of the hypercharge.
\begin{table}
\begin{center}
\begin{tabular}{|c|c|} \hline
  Initial state & Final states \\ \hline
  $\tilde{\chi} \tilde{\chi}$ & $f \bar{f}$ \\ \hline
  $\tilde{l}_{R}^{i} \tilde{\chi}$ & $l^{i}\gamma, l^{i} Z, l^{i} h$ \\ \hline
  $\tilde{l}_{R}^{i} \tilde{l}_{R}^{i^{*}}$ & $\gamma \gamma, ZZ, 
      \gamma Z, W^{+}W^{-}, Zh, \gamma h, hh, f \bar{f}$ \\ \hline
  $\tilde{l}_{R}^{i} \tilde{l}_{R}^{j}$ & $l^{i} l^{j}$ \\ \hline
  $\tilde{l}_{R}^{i} \tilde{l}_{R}^{j^{*}}  (i \neq j)$ & $l^{i} 
      \bar{l}^{j}$ \\ \hline
\end{tabular}
\end{center}
\caption{Annihilation cross sections; $i,j$ = $\tau,\mu,e$.}
\label{tab.coanh}
\end{table}

To estimate the relic LSP abundance, we need $\vev{\sigma_{\rm eff}v}$
at the freeze-out temperature $T_{f}$ of the LSP.  Since typically 
$T_{f} \sim m_{\bino}/25$ where $m_{\bino}$ is the LSP mass, the 
expansion of $\vev{\sigma_{\rm eff}v}$ in terms of $T/m_{\tilde{\chi}}$ 
(partial-wave expansion) is relevant.  Thermally-averaged cross section 
$\vev{\sigma_{ij}v}$ for the process $i+j \ra k+l$ is given by \cite{EFOS}:
\begin{equation}
  \vev{\sigma_{ij}v} = \frac{1}{m_{i}m_{j}} \left[
    1 - \frac{3(m_{i}+m_{j})}{2 m_{i}m_{j}} T \right] \: 
    w(s) |_{s \ra (m_{i}+m_{j})^{2} + 3(m_{i}+m_{j})T} \: 
    + {\cal O}\left( \frac{ T^{2} }{m_{\tilde{\chi}}^2 }\right),
\end{equation}
where
\begin{equation}
  w(s) \equiv \frac{1}{4} \int \frac{d^{3}p_{k}}{(2\pi)^{3} E_{k}}
    \frac{d^{3}p_{l}}{(2\pi)^{3} E_{l}} \: (2\pi)^{4} 
    \delta^{4}(p_{i}+p_{j}-p_{k}-p_{l}) |{\cal T}|^{2}.
\end{equation}
Here, $s = (p_{i}+ p_{j})^{2}$ is the Mandelstam variable, and 
$|{\cal T}|^{2}$ is the transition matrix element squared summed over 
final state spins and averaged over initial state spins.  Terms of order
$(T/m_{\bino})^{0}$ and $(T/m_{\bino})^{1}$ are called $s$-wave
and $p$-wave components, respectively.

The $s$-wave component of the thermally-averaged neutralino annihilation
cross section $\vev{\sigma_{\tilde{\chi} \tilde{\chi}}v}$ is suppressed
by tiny final-state fermion masses, so that the $p$-wave part is 
the dominant piece. 
While neutralino annihilation cross section is $p$-wave suppressed, 
those for the sleptons, $\vev{\sigma_{\tilde{l}_{i} \tilde{l_{j}} } v}$, 
have $s$-wave components as dominant pieces.  Therefore, if the 
equilibrium number densities for the sleptons are not much smaller than 
that for the neutralino, the slepton annihilation processes can 
significantly reduce the relic LSP abundance (see Eq.~(\ref{sigma_eff})).
This coannihilation process is effective when the slepton masses are 
degenerate with the neutralino mass within $\sim 20\%$.
In the GMSB spectrum, it happens when $\tan \beta$ is large and/or 
$N$ is greater than 1.

The partial-wave expansion of the thermally-averaged 
cross section does not give a good approximation when the initial momentum 
is near the $s$-channel pole or the final-state threshold \cite{GS}.
They could occur at $Z,h,A,H$ poles and $WW, ZZ, Zh, t \bar{t}$ thresholds 
in the neutralino annihilation.  However, with the present GMSB mass 
spectrum, these cases simply do not happen or their effects are strongly 
suppressed.  First, the $Z$, $h$ poles can be hit when 
$m_{\tilde{\chi}} \sim m_{Z}/2, m_{h}/2$, but these regions are already 
being excluded by the chargino search at LEP2.  The situations 
$m_{\tilde{\chi}} \sim m_{A}/2, m_{H}/2$ also do not occur in the 
parameter region we are considering.  As for the threshold effect, 
$WW$ and $ZZ$ ones are small due to $\mu \gg M_1$; $Zh$ one is also 
negligible due to the smallness of the Higgsino component in 
$\tilde{\chi}$, and $t \bar{t}$ one is strongly suppressed by 
the large masses for the top squarks exchanged in $t$-channel.

\begin{figure}[ht]
\begin{center}
\rotatebox[origin=c]{-90}{
\includegraphics[width=11cm]{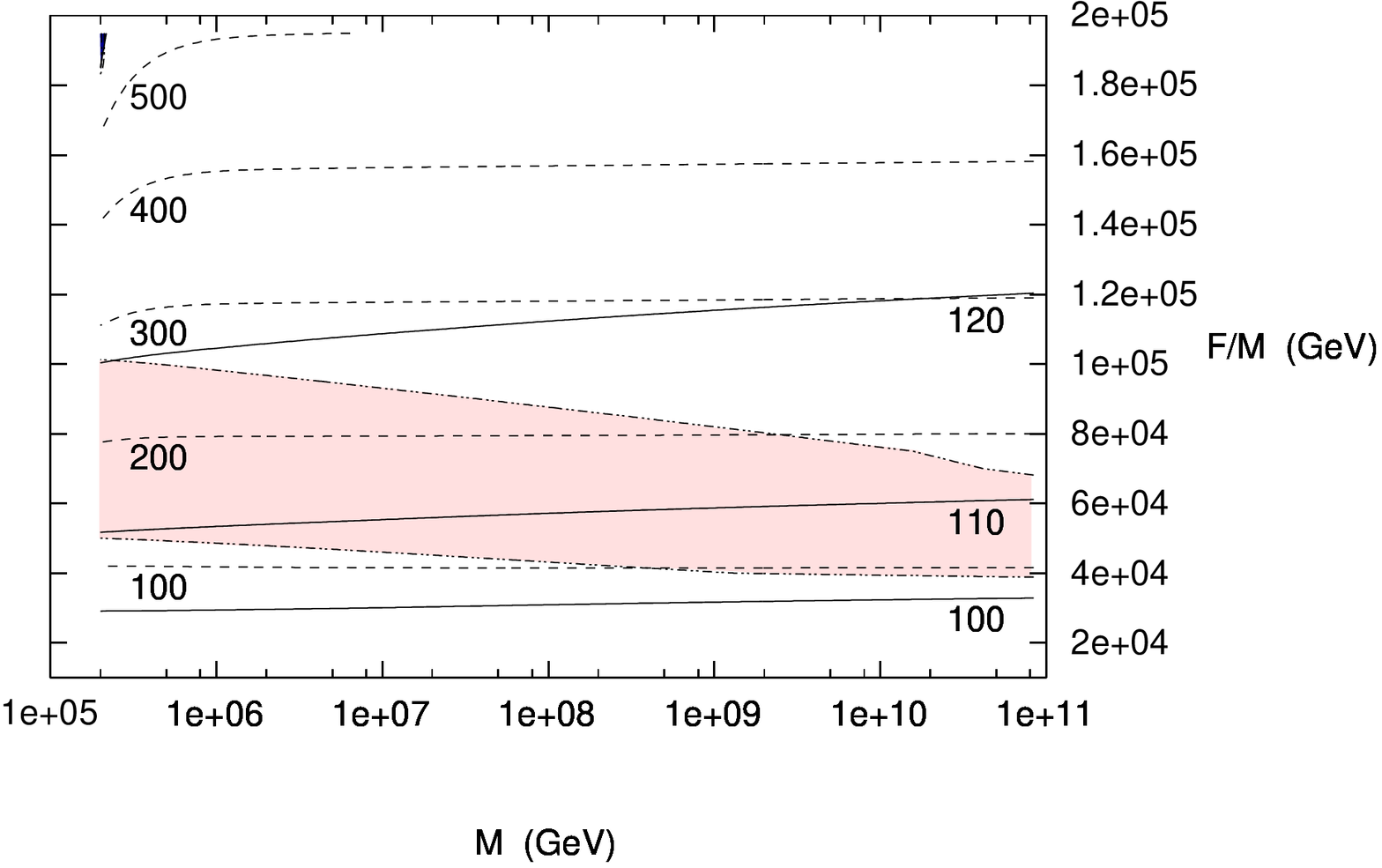}}
\end{center}
\caption{GMSB parameter space ($M, F/M$) corresponding to
 $0.1 \lsim \Omega_{\tilde{\chi}} h^{2} \lsim 0.3$ (the light shaded
 region) for $\tan\beta = 10$, sgn$\mu = +1$. The dashed lines 100, 200,
 300, 400, 500 are the contours for the lightest chargino mass, 
 and the solid lines 100, 110, 120 are the contours for the Higgs 
 boson mass (in unit of GeV). The dark shaded region in the
 upper left corner is the stau LSP region.}
\label{fig:N1-tan10-1}
\end{figure}

With these understandings, we can calculate the relic abundance 
$\Omega_{\tilde{\chi}} h^2$.  In the actual calculation, we have used 
the computer program {\it neutdriver} coded by Jungman, Kamionkowski 
and Griest \cite{DM_hidden-3}, which contains all the (co)annihilation
cross sections calculated by Drees and Nojiri \cite{DM_hidden-2}.
In Fig.~\ref{fig:N1-tan10-1}, we have shown a cosmologically favored 
region, $0.1 < \Omega_{\tilde{\chi}} h^{2} < 0.3$ (light shaded
regions), on $M$--$F/M$ plane in the case of  $(N,\tan\beta) = (1,10)$.
We scanned the region $10^4~{\rm GeV} \lsim F/M \lsim 2 \times 
10^5~{\rm GeV}$, which corresponds to the soft SUSY-breaking masses 
$\sim 100~{\rm GeV}$ -- $1~{\rm TeV}$.  The sign of $\mu$ is taken to be 
positive in the standard notation (the one in which the constraint from 
$b \rightarrow s \gamma$ process is weaker). The region extends from
upper left to lower right directions.  This is because the relic
abundance is almost completely determined by the mass of the
right-handed stau, which is monotonically increasing with $M$ with a 
fixed value of $F/M$ due to renormalization group effects.  We have 
also drawn the contours of the lightest Higgs boson mass (solid lines) 
and the lightest chargino mass (dashed lines) in GeV, which are 
calculated using {\it neutdriver}.  We find that some of the parameter 
region satisfies constraints from the lower bounds on the Higgs boson 
mass $m_{h^0} \gsim 113.5~{\rm GeV}$ and the lightest chargino mass 
$m_{\tilde{\chi}^{\pm}} \gsim 150~{\rm GeV}$.  In particular, we find
that smaller messenger scale, $M \lsim 10^8~{\rm GeV}$, is favored.

\begin{figure}[ht]
\begin{center}
\rotatebox[origin=c]{-90}{
\includegraphics[width=11cm]{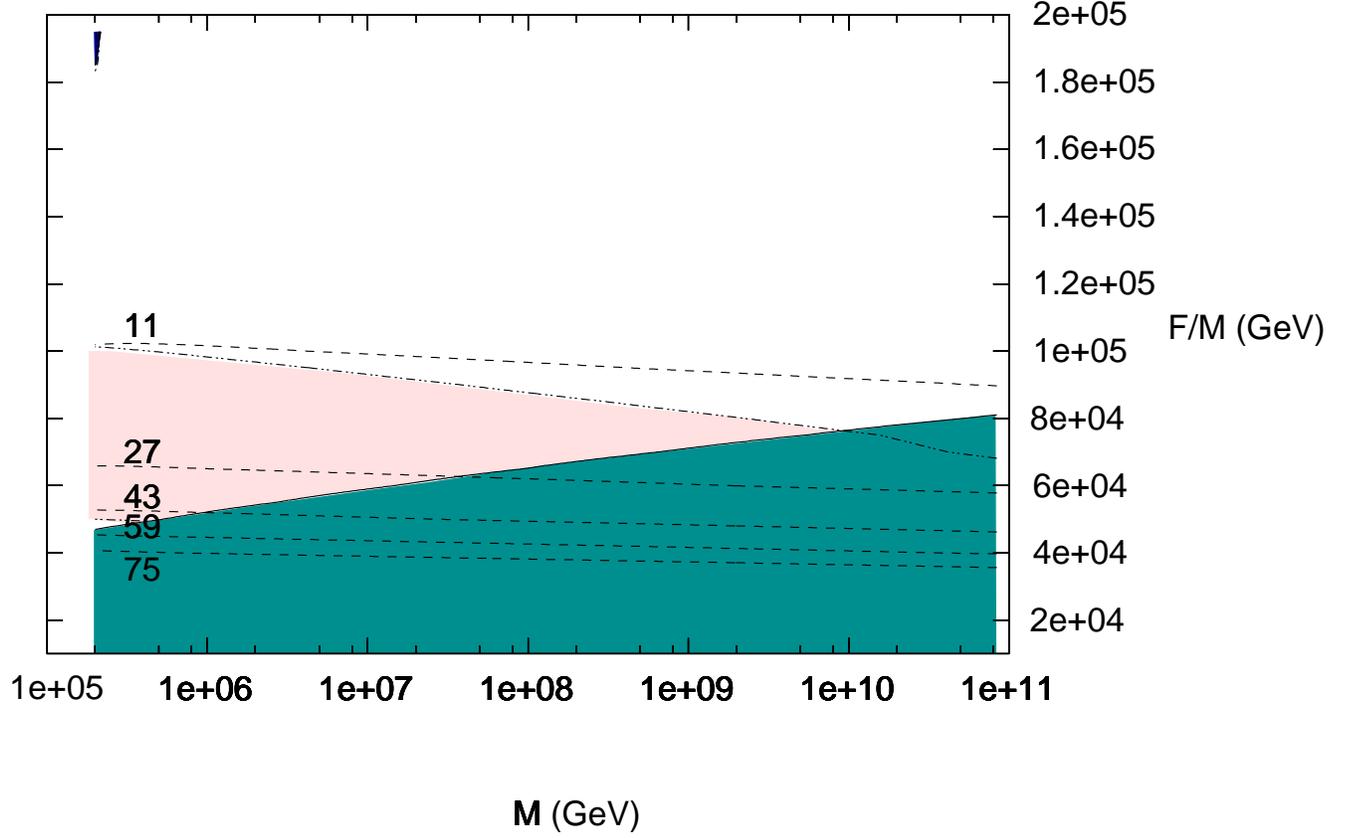}}
\end{center}
\caption{GMSB parameter space ($M, F/M$) corresponding to $0.1
 \lsim \Omega_{\tilde{\chi}} h^{2} \lsim 0.3$ (the light shaded region)
 for $\tan\beta = 10$, sgn$\mu = +1$.
 The dotted lines 11, 27, 43, 59, 75 are the contours of the $-2\sigma$,
 $-1\sigma$, central, $+1\sigma$, and $+2\sigma$ values of the 
 SUSY contribution to the muon ($g-2$), $a_{\mu}$, in unit of $10^{-10}$.
 The dark shaded region in the bottom is excluded by $b \ra s \gamma$.} 
\label{fig:N1-tan10-2}
\end{figure}

In Fig.~\ref{fig:N1-tan10-2}, we have shown the constraint from 
$b \ra s \gamma$, $2.3 \times 10^{-4} < Br(b \ra s \gamma) < 
4.1 \times 10^{-4}$ \cite{CLEO}, in the same $M$--$F/M$ plane as
Fig.~\ref{fig:N1-tan10-1}.  The dark shaded region indicates the
excluded region.  We see that most of the parameter region which 
satisfies the constraints from the Higgs and chargino masses also 
satisfies that from $b \ra s \gamma$.  We have also drawn the contour 
of the SUSY contribution to the muon anomalous magnetic moment, 
$a_{\mu}$, in units of $10^{-10}$.  The SUSY contributions to 
$a_{\mu}$ were discussed, for example, in Refs.~\cite{muon-papers-1, 
muon-papers-2, muon-papers-gmsb, muon-papers-recent}.  It is 
interesting that the value is consistent with that required to 
explain the recently claimed 2.6$\sigma$ deviation of $a_{\mu}$ between 
the observed and the standard-model values.

\begin{figure}[ht]
\begin{center}
\rotatebox[origin=c]{-90}{
\includegraphics[width=11cm]{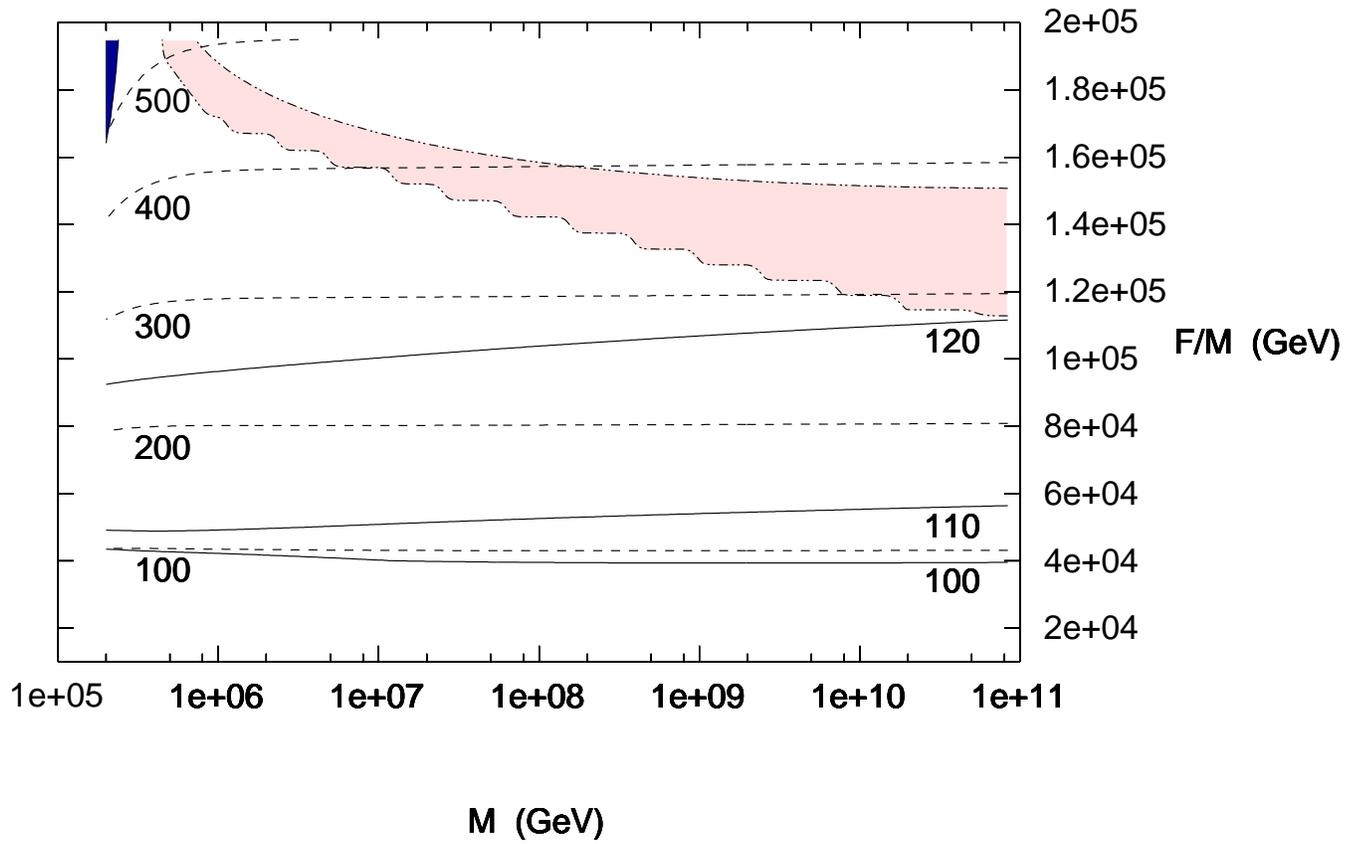}}
\end{center}
\caption{The same figure as Fig.~\ref{fig:N1-tan10-1} for $\tan\beta = 50$.}
\label{fig:N1-tan50-1}
\end{figure}

\begin{figure}[ht]
\begin{center}
\rotatebox[origin=c]{-90}{
\includegraphics[width=11cm]{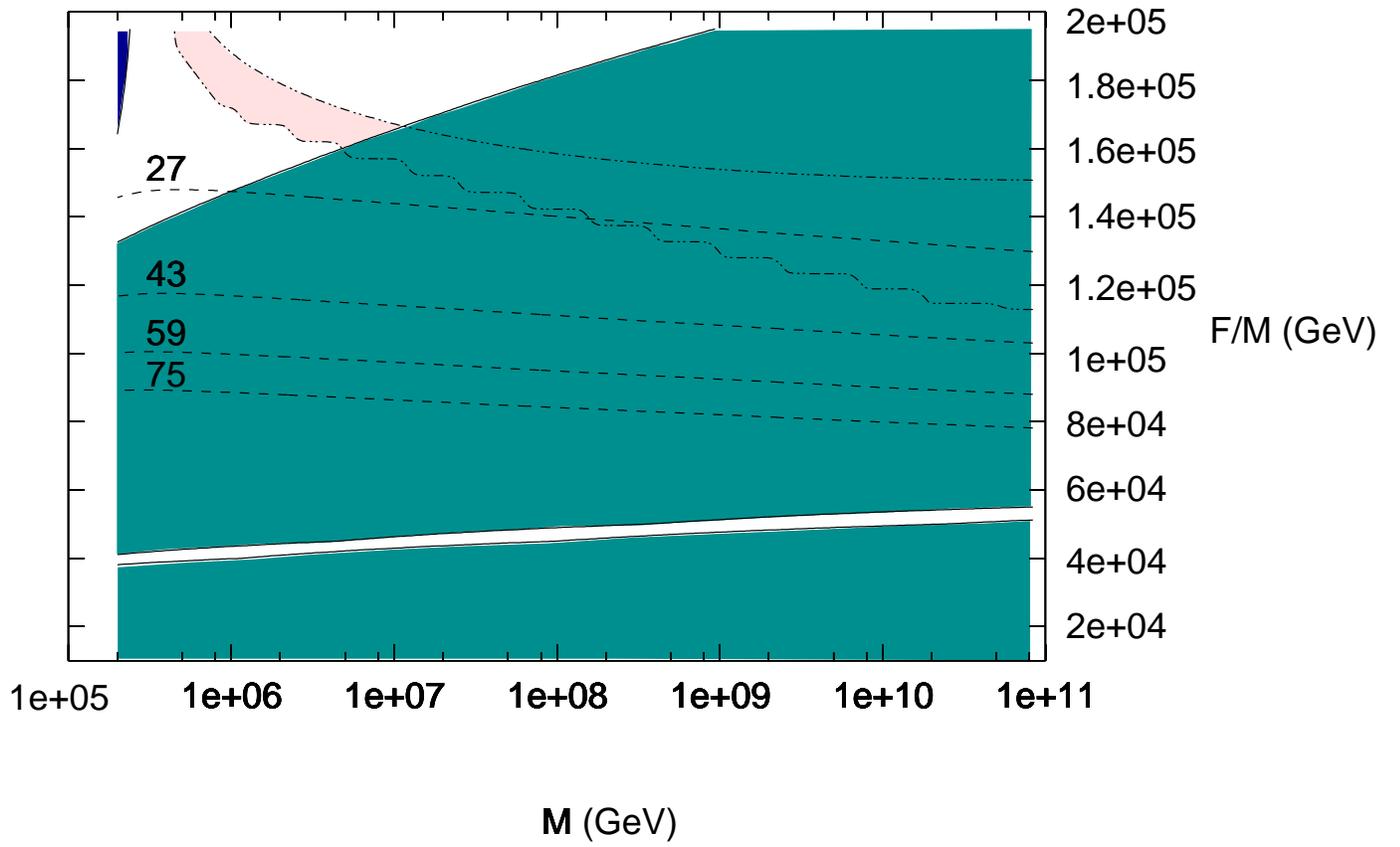}}
\end{center}
\caption{The same figure as Fig.~\ref{fig:N1-tan10-2} for $\tan\beta = 50$.
 The narrow strip in the dark shaded region (excluded by $b\ra s \gamma$)
 is an accidentally allowed region.} 
\label{fig:N1-tan50-2}
\end{figure}

In Figs.~\ref{fig:N1-tan50-1} and \ref{fig:N1-tan50-2}, we have plotted 
various quantities in the case of $(N,\tan\beta) = (1,50)$.
In this case, the coannihilation effect is important so that the
cosmologically favored region (light shaded region) is shifted to 
larger values of $F/M$ in lower $M$ region ($M \sim F/M$).
We find that the constraint from $b \ra s \gamma$ is more stringent 
than the $\tan\beta = 10$ case, but the region $M \lsim 10^7~{\rm GeV}$ 
still satisfies the constraints.  This region gives the value of 
$a_{\mu}|_{\rm SUSY}$ within $1 \sigma$ and $2 \sigma$ level for the
SUSY explanation of the observed deviation.

\begin{figure}[ht]
\begin{center}
\rotatebox[origin=c]{-90}{
\includegraphics[width=11cm]{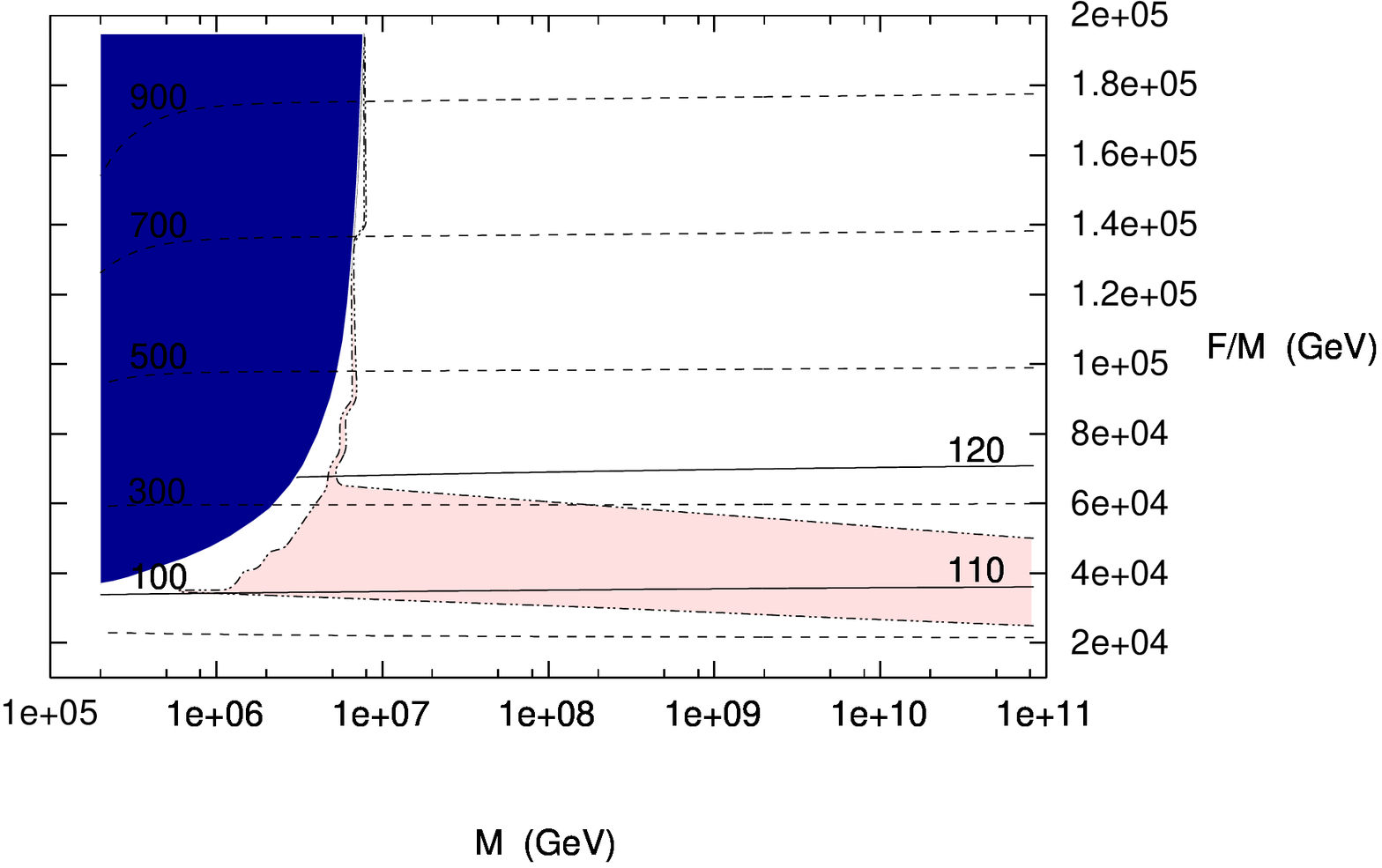}}
\end{center}
\caption{The same figure as Fig.~\ref{fig:N1-tan10-1} for $N=2$. The dark 
 shaded region in the upper left (stau LSP region) is larger than that
 for $N=1$.}
\label{fig:N2-tan10-1}
\end{figure}

\begin{figure}[ht]
\begin{center}
\rotatebox[origin=c]{-90}{
\includegraphics[width=11cm]{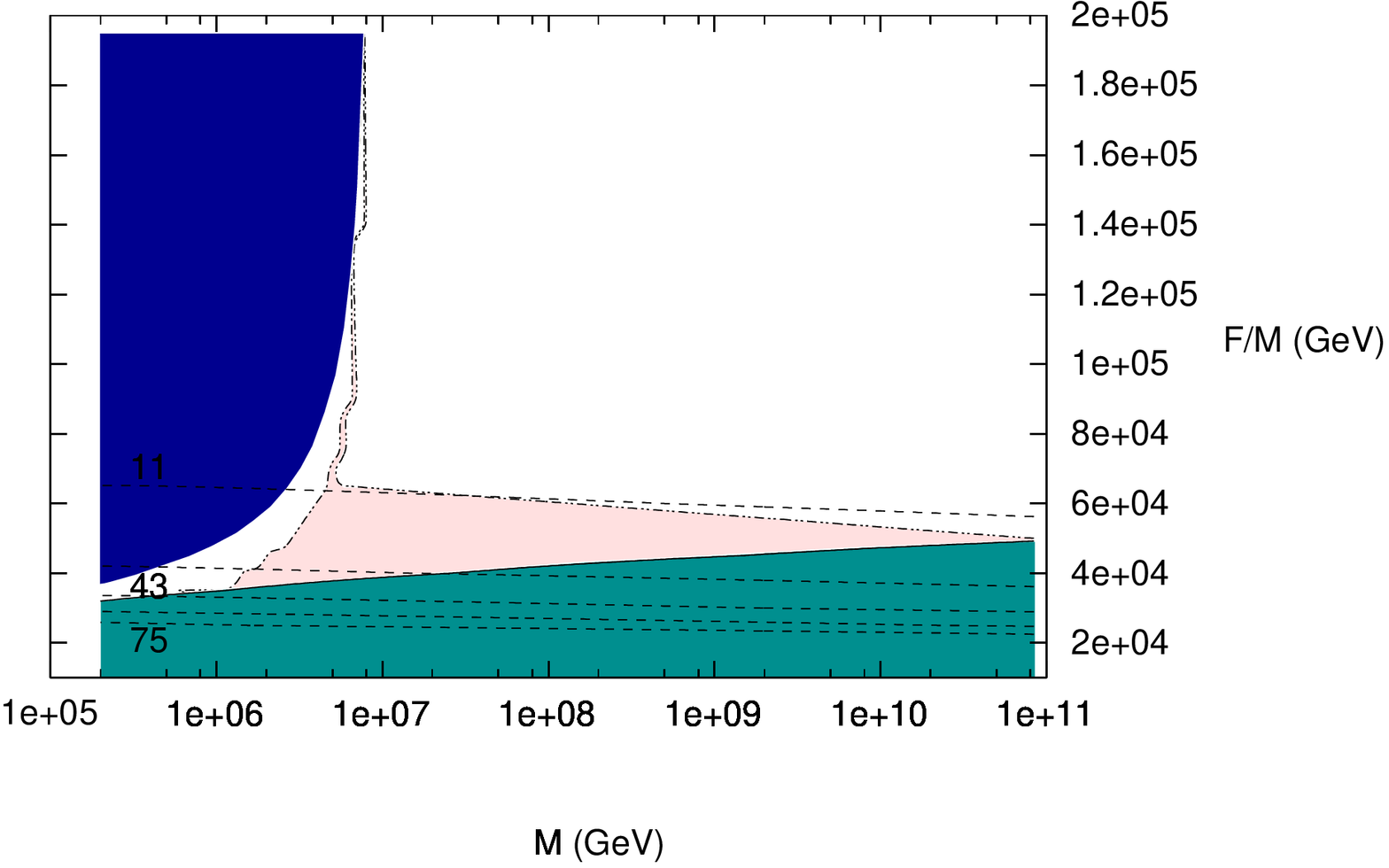}}
\end{center}
\caption{The same figure as Fig.~\ref{fig:N1-tan10-2} for $N=2$. The dark 
 shaded region in the upper left (stau LSP region) is larger than that
 for $N=1$.} 
\label{fig:N2-tan10-2}
\end{figure}

\begin{figure}[ht]
\begin{center}
\rotatebox[origin=c]{-90}{
\includegraphics[width=11cm]{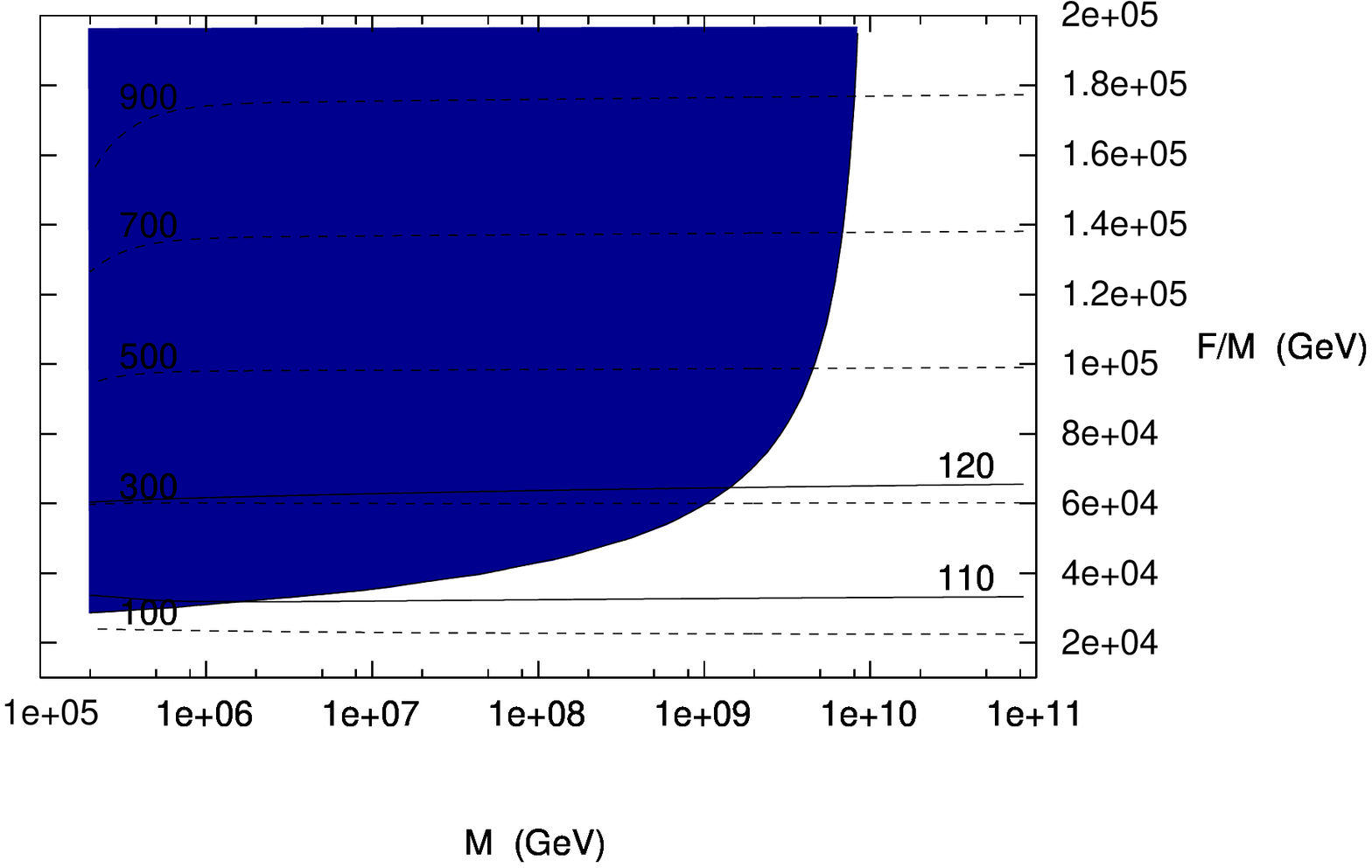}}
\end{center}
\caption{The same figure as Fig.~\ref{fig:N1-tan50-1} for $N=2$. In this
 case the region for $0.1 \lsim \Omega_{\tilde{\chi}} h^{2} \lsim 0.3$
 is absent in the range of the parameters exhibited.} 
\label{fig:N2-tan50-1}
\end{figure}

\begin{figure}[ht]
\begin{center}
\rotatebox[origin=c]{-90}{
\includegraphics[width=11cm]{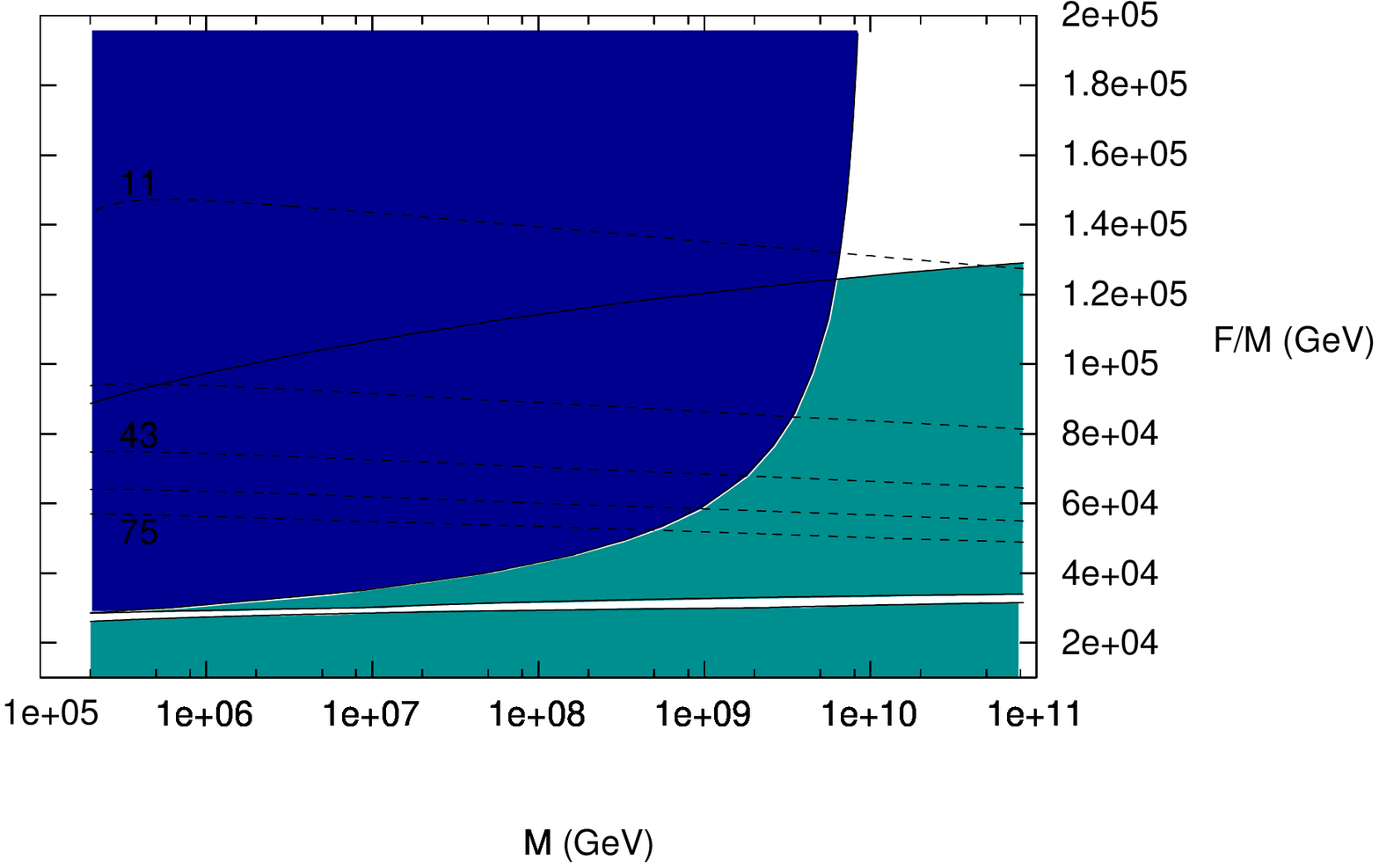}}
\end{center}
\caption{The same figure as Fig.~\ref{fig:N1-tan50-2} for $N=2$. In this
 case the region for $0.1 \lsim \Omega_{\tilde{\chi}} h^{2} \lsim 0.3$
 is absent in the range of the parameters exhibited.} 
\label{fig:N2-tan50-2}
\end{figure}

We have also done the same analyses in the case of $N=2$, which are
plotted in Figs.~\ref{fig:N2-tan10-1}, \ref{fig:N2-tan10-2}, 
\ref{fig:N2-tan50-1} and \ref{fig:N2-tan50-2}.  The black region
represents the region where the stau is the lightest SUSY particle.
In the case of $\tan\beta = 10$ there is a consistent region which
reproduces cosmologically interesting abundance of the LSP, but 
we do not find such a region in the case of $\tan\beta = 50$.

\section{Conclusions and Discussion}

In this paper we have investigated gauge mediation models with 
neutralino LSP.  This type of models is naturally realized in 
the extra-dimensional setup \cite{NY}.  Due to the geometrical separation 
between the SUSY-breaking and observable sectors, flavor non-universal 
contributions to the squark and slepton masses are exponentially 
suppressed, which makes it possible to solve the $\mu$ problem by the 
mechanism of Ref.~\cite{GM}.  The consequence is that the gravitino 
mass becomes of the order of the weak scale, while all the other 
superparticle masses are the same as those in usual GMSB models.
Therefore, the lightest neutralino (mostly bino) could provide 
the cold dark matter of the universe.

We have calculated the relic abundance of the LSP neutralino in 
this brane-world GMSB model.  We found that there is a wide parameter 
region where cosmologically favored relic densities, 
$0.1 < \Omega_{\tilde{\chi}} h^{2} < 0.3$, are obtained and, at the same 
time, the lower bounds on the Higgs and the lightest chargino masses from 
collider experiments are evaded.  We have also found that the experimental
constraint on the $b \rightarrow s\gamma$ process is satisfied in this
cosmologically favored parameter region, 
especially when the messenger scale $M$ is low.  Interestingly, this 
parameter region also gives the SUSY contribution to the muon anomalous 
magnetic moment, $a_{\mu}|_{\rm SUSY}$, with the values required to 
explain the recently reported 2.6$\sigma$ deviation of $a_{\mu}$ 
\cite{muon} from the standard-model value.  These are due to the fact that 
in the GMSB spectrum the colored superparticles are relatively heavy while 
non-colored ones are light; the constraint from $b \rightarrow s\gamma$ 
is easily evaded since squarks are heavy, while we obtain relatively large 
values of $a_{\mu}|_{\rm SUSY}$ since sleptons are light.  Although 
we have limited ourselves to the case of the minimal messenger sector 
in this paper, the above features are generic to GMSB models, for example, 
ones with more general messenger sectors \cite{SP-M, R-CRS, INTY}.  
We leave the study of the LSP relic abundance under these more general 
superparticle spectrums for future work.

It is interesting to compare the present situation with other SUSY-breaking 
models that solve the SUSY flavor problem.  In anomaly mediated 
SUSY-breaking scenario \cite{RS, GLMR}, the LSP is generically wino-like 
neutralino, and its annihilation cross section is too large to obtain 
cosmologically favored thermal relic density of the LSP, 
so that we need to consider some other source 
of the LSP, such as the decay of the moduli field \cite{MR}.
In the decoupling scenario \cite{decoupling} and the focus-point 
scenario \cite{focus}, it would be difficult to accommodate sufficiently 
large values of the muon anomalous magnetic moment since the smuons 
are assumed to be heavy in these scenarios.  The gaugino mediation 
scenario \cite{gaugino} is a promising candidate theory, but the 
parameter region of the neutralino LSP is not necessarily very large, 
especially when the Higgs fields are localized on the observable brane.
Thus, we conclude that the GMSB models with neutralino dark matter 
provide one of the promising SUSY-breaking scenarios which are 
in accordance with current phenomenological situations.  It would be 
interesting to study further the relic abundance of the LSP neutralino 
and its detections in this context.

\section*{Acknowledgments}

The work of Y.N. was supported by the Miller Institute for 
Basic Research in Science. K.S. thanks the Japan Society 
for the Promotion of Science for financial support.

\newpage

%
%
\newcommand{\Journal}[4]{{\sl #1} {\bf #2} {(#3)} {#4}}
\newcommand{\PL}{\sl Phys. Lett.}
\newcommand{\PR}{\sl Phys. Rev.}
\newcommand{\PRL}{\sl Phys. Rev. Lett.}
\newcommand{\NP}{\sl Nucl. Phys.}
\newcommand{\ZP}{\sl Z. Phys.}
\newcommand{\PTP}{\sl Prog. Theor. Phys.}
\newcommand{\NC}{\sl Nuovo Cimento}
\newcommand{\MPL}{\sl Mod. Phys. Lett.}
\newcommand{\PRep}{\sl Phys. Rept.}

%
\newpage

\end{document}